\documentclass[runningheads]{llncs}
\usepackage[T1]{fontenc}
\usepackage[misc,geometry]{ifsym}

\usepackage{graphicx, amsmath, amssymb, amsfonts}
\usepackage{array, ragged2e, vcell, tabularray, diagbox}
\usepackage{subcaption}
\usepackage{lipsum}
\usepackage[hidelinks]{hyperref}
\usepackage{color}
\usepackage{tabularray}

\usepackage{float}

\usepackage[acronym]{glossaries}

\makeglossaries

\newacronym{ml}{ML}{machine learning}
\newacronym{6g}{6G}{$6^\text{th}$ Generation Technology}
\newacronym{elaa}{ELAA}{Extreme Large Antenna Array}
\newacronym{nfc}{NFC}{near-field communication}
\newacronym{ffc}{FFC}{far-field communication}
\newacronym{pw}{PW}{planar wavefront}
\newacronym{sw}{SW}{spherical wavefront}
\newacronym{racnn}{RACNN}{Residual Attention and Convolution Neural Networking}
\newacronym{xlcnet}{XLCNet}{XLCNet}
\newacronym{dl}{DL}{deep learning}
\newacronym{ls}{LS}{Least Square}
\newacronym{mmse}{MMSE}{Minimum Mean Square Error}
\newacronym{nmse}{NMSE}{Normalized Mean Square Error}
\newacronym{ff}{FF}{far-field}
\newacronym{nf}{NF}{near-field}
\newacronym{hf}{HF}{hybrid-field}
\newacronym{cnn}{CNN}{Convolutional Neural Network}
\newacronym{ra}{RA}{Residual Attention}
\newacronym{fpn}{FPN}{Fixed Point Networks}
\newacronym{bs}{BS}{base station}
\newacronym{ue}{UE}{user equipment}
\newacronym{em}{EM}{electromagnet}

\begin{document}
\title{RACNN: Residual Attention Convolutional Neural Network for Near-Field Channel Estimation in 6G Wireless Communications}
\titlerunning{RACNN: Near-field Channel Estimation for 6G}
%
\author{Vu Tung Lam\inst{1} \and Do Hai Son\inst{2} \and Tran Thi Thuy Quynh\inst{1}  {\small \Letter} 
\and Thanh Trung Le \inst{1} 
}

\authorrunning{Vu Tung Lam et al.}
%
\institute{VNU University of Engineering and Technology, VNU Hanoi, Vietnam \and
VNU Information Technology Institute, VNU Hanoi, Vietnam \\
    \email{Emails: \big\{21025112, dohaison1998, quynhttt, thanhletrung\big\}@vnu.edu.vn} 
}
\maketitle              
%

\setcounter{footnote}{0}

\begin{abstract}

Near-field channel estimation is a fundamental challenge in the sixth-generation (6G) wireless communication, where extremely large antenna arrays (ELAA) enable near-field communication (NFC) but introduce significant signal processing complexity. Traditional model-based methods suffer from high computational costs and limited scalability in large-scale ELAA systems, while existing learning-based approaches often lack robustness across diverse channel conditions. To overcome these limitations, we propose the Residual Attention Convolutional Neural Network (RACNN), which integrates convolutional layers with self-attention mechanisms to enhance feature extraction by focusing on key regions within the CNN feature maps. Experimental results show that RACNN outperforms both traditional and learning-based methods, including XLCNet, across various scenarios, particularly in mixed far-field and near-field conditions. Notably, in these challenging settings, RACNN achieves a normalized mean square error (NMSE) of $4.8*10^{-3}$ at an SNR of 20dB, making it a promising solution for near-field channel estimation in 6G.
 
\keywords{Near-field communication (NFC)  \and 6G \and Channel estimation \and Deep learning  \and ELAA}
\end{abstract}

\section{Introduction}\label{sec:intro}

Sixth-generation technology (6G) is expected to be the next major advancement in telecommunications. The key performance indicators of 6G should be much superior to those of 5G in terms of peak data rate, spectral efficiency, access density, coverage, and latency~\cite{Jiang2021}. To achieve these indicators, 6G requires a significantly increased number of antennas, i.e., thousands of elements, compared to hundreds of elements in 5G. Antenna arrays that meet this requirement are referred to as extremely large antenna arrays (ELAA), distinguishing them from massive multiple-input multiple-output (MIMO) in 5G.

ELAA is not only more complex in design and deployment but also introduces new challenges in signal processing~\cite{Wei2022}, particularly the fundamental transition in electromagnetic propagation modeling from planar-wave (i.e., far-field communication in 5G) to spherical-wave (i.e., near-field communication in 6G). In detail, the Rayleigh distance is used to determine the boundary where propagation is divided as near-field or far-field. For instance,~\cite{cui2022near,Wei2022} indicate that wireless communication systems operating with ELAA and within the Terahertz frequency range (typically around 300GHz) can extend the Rayleigh distance beyond 200~meters, often covering a significant portion of user equipment (UE) devices. This implies that most UEs will operate within the near-field propagation region. Hence, the near-field communication in 6G (NFC-6G) introduces specifically two major challenges: (i) the substantial computational complexity as the number of ELAA elements reaches thousands; (ii) all existing signal processing techniques used in 5G and earlier generations must be adjusted to work with the near-field channel model. In this study, we focus on the problem of channel~estimation.

Channel estimation is an essential problem in wireless communication systems and many studies have been proposed to address this problem in far-field scenarios~\cite{Hassan2020}. These studies can be divided into two groups: (i) traditional model-based methods, such as least squares (LS) and minimum mean square error~(MMSE)~\cite{Kay1993}, and (ii) learning-based methods, which have recently gained attention for leveraging machine learning techniques in channel estimation. Methods in the former group are simple to implement in practice. However, their estimation accuracy is affected by noise and interference. Moreover, when the number of ELAA elements grows larger, the matrix inversion required in these methods is computationally challenging. The latter group is further divided into two approaches, i.e., model-driven methods and data-driven methods. The model-driven approach focuses on optimizing the model based on mathematical solutions. For example, in~\cite{Zhang2023}, the authors presented learning ISTA (LISTA) and sparsifying dictionary learning-LISTA (SDL-LISTA) networks, which are neural networks based on the iterative shrinkage and thresholding algorithm~(ISTA). In~\cite{qiang2020}, the authors proposed AMP-NET based on the principle of approximate message passing (AMP) for joint activity detection and channel estimation. These methods achieve higher estimation accuracy compared to the original methods with a small number of learning parameters. However, a neural network with a small number of learning parameters also has its own drawbacks. They perform well only under specific channel conditions that match their training data, such as the number of far-field paths, near-field paths, and noise distribution. The data-driven approach can mitigate this limitation by training neural networks with a large number of learning parameters. This enables the networks to learn from a wide range of channel conditions. To illustrate, in~\cite{gao2024}, the authors proposed an extremely large-scale MIMO channel network (XLCNet) for channel estimation in NFC-6G systems. Specifically, XLCNet transforms noisy channel vectors into 2-channel images and then uses a convolutional neural network (CNN) to denoise them. Notably, this network supports not only far-field and near-field scenarios but also mixed far-field/near-field scenarios. Moreover, CNN is a widely studied architecture that has been utilized for many years. With recent advancements in learning-based techniques for image processing and natural language processing, the underlying concept of XLCNet has the potential for further enhancement.

Inspired by~\cite{gao2024} and a recent advancement in deep learning (DL) known as the attention mechanism~\cite{Vaswani2017}, we propose a novel network named the RACNN for NFC-6G wireless communication.\footnote{Simulation codes: \url{https://github.com/DoHaiSon/RACNN}.} RACNN inherits the idea of estimating noise through a neural network and subtracting the estimated noise from the input at the final stage. RACNN follows the convolutional block structure of XLCNet and incorporates self-attention blocks to enhance the overall performance of noise estimation. Our numerical results show that RACNN outperforms traditional model-based methods (LS and MMSE) and XLCNet in various scenarios, e.g., 
(i) matched training and testing configurations in terms of far-/near-field paths; (ii) testing with only far-field or only near-field paths; and (iii) mismatched mixtures of far-field and near-field components.

\section{System Model}\label{sec:sm}

\subsection{Rayleigh Distance}
\label{subsec:Rayleigh distance}

The Rayleigh distance of an NFC-6G system can be represented as follows
\begin{equation}
    D_{\text{Ray}} = \frac{2(M\frac{\lambda}{2})^2}{\lambda}=\frac{1}{2}M^2\lambda,
    \label{eq:Rayleigh_distance} 
\end{equation} where $M$ is the array dimension and $\lambda$ is the signal wavelength~\cite{Sun2025}. 
In NFC-6G systems, the antenna array can reach thousands of elements, the transmit wavelength is shorter, and the aperture is larger. As a result, the Rayleigh distance $D_{\text{Ray}}$ increases, extending to typical operating distances in cellular networks. This shift places UE services within the near-field range~\cite{Sun2025}. Unlike far-field channels, where signals exhibit planar wavefronts, signals in near-field channels propagate with spherical wavefronts and spread-out patterns, resulting in non-uniform distribution characteristics at the receiver. To understand far-field and near-field physical properties, we analyze them over mathematical channel functions presented in the next subsection.

\subsection{Channel Models}
We consider an NFC-6G wireless channel with an ELAA consisting of $M$ elements, serving $K$ UEs. Following the article in~\cite{gao2024}, the downlink far-field channel is modeled as
\begin{equation}
    \mathbf{Y} = \mathbf{H}\mathbf{P}^{\frac{1}{2}}\mathbf{S}+\mathbf{Z},
    \label{eq:signal_transmission_model}
\end{equation}
where $\mathbf{H} \in \mathbb{C}^{M \times K}$ represents the channel matrix, $\mathbf{P} = \operatorname{diag} \left(\mathit{P}_1, \mathit{P}_2, \dots  ,\mathit{P}_K\right)$ is the transmit
power matrix, $\mathbf{S} \in \mathbb{C}^{K\times K}$ is the transmit pilot matrix, and $\mathbf{Z}~\in~\mathbb{C}^{M\times K}$ is an additive white Gaussian noise matrix. We also assume that signal transmission is lossless and transmit powers across all antennas are identical, i.e., $P_k = P ~\forall k$. In NFC-6G systems, the channel matrix $\mathbf{H}$ can be expressed in a unified form for both far-field and near-field cases~\cite{Wei2022} as follows:
\begin{equation}
    \mathbf{H} =\mathbf{A}\mathbf{G},
    \label{eq:channel_matrix_unified}
\end{equation}
where $\mathbf{G}\in\mathbb{C}^{M\times L}$ is a gain matrix whose elements are independent identically distributed from $\mathcal{CN}(0,\sigma^2)$, $\mathbf{A}\in\mathbb{C}^{M\times L}$ is the steering matrix (which can be the far-field $\mathbf{A}_\text{far}$,  near-field $\mathbf{A}_\text{near}$, and hybrid $\mathbf{A}_\text{hybrid}$), and $L$ is the number of multipaths \cite{Wei2022}. Particularly, the steering matrix $\mathbf{A}_\text{far}$ in far-field cases is given~by   
\begin{equation}
    \mathbf{A}_\text{far}=\sqrt{\frac{M}{L}}\big[\mathbf{a}(\phi_1), \mathbf{a}(\phi_2), \cdots,\mathbf{a}(\phi_l)\big].
    \label{eq:far_field_steering_matrix}
\end{equation}
Here, consider a uniform linear array (ULA) at the BS, where each channel vector $\mathbf{a}(\phi_l)$ is expressed as
\begin{equation}
        \mathbf{a}(\phi_l) = \frac{1}{\sqrt{M}}\left[e^{-j2\pi\frac{d}{\lambda}\sin(\phi_l)},\dots, e^{-j2\pi\frac{d}{\lambda}(M-1)\sin(\phi_l)}\right]^\top,
    \label{eq:far_field_steering_vectore}
\end{equation}
where $d = \frac{\lambda}{2}$ denotes the antenna spacing and $\phi_l= \left[-\frac{\pi}{2};\frac{\pi}{2}\right]$ be the steering angle of the $l$-th path. On the other hand, the steering matrix  $\mathbf{A}_\text{near}$ in near-field case is parameterized by $\phi_l$ and $r_l$ as follows
\begin{equation}
    \mathbf{A}_\text{near} = \sqrt{\frac{M}{L}}\left[\mathbf{a}(\phi_1,r_1),\cdots,\mathbf{a}(\phi_L,r_L)\right],\,
    \label{eq:near_field_steering_matrix}
\end{equation}
with the steering vector $\mathbf{a}(\phi_l,r_l)$ is defined as 
\begin{equation}
    \mathbf{a}(\phi_l,r_l) = \frac{1}{\sqrt{M}}\left[e^{-j\frac{2\pi}{\lambda}(r_{l,1}-r_l)},\cdots,e^{-j\frac{2\pi}{\lambda}(r_{l,M}-r_l)}\right]^\top,
    \label{eq:near_field_steering_vector}
\end{equation}
where $r_l$ represents the distance from the $l$-th scatter to the center of the antenna array, while  $r_{l,m} = \sqrt{r^2_l + \delta^2_m d^2 - 2 r_l \delta_m d \sin(\phi_l)}$ is the distance between the \mbox{$l$-th} scatter and the $m$-th antenna element at the base station (BS) with $\delta_m = \frac{2m - M -1}{2}$ and $m = 1,2, \cdots, M$. 

Consider that the gain $\mathbf{G}$ in NFC-6G systems remains similar when signals propagate among far-field and near-field regions. Accordingly, the \acrlong{hf} channel model $\mathbf{H}_\text{hybrid}$  adopts the following form
\begin{equation}
    \mathbf{H}_\text{hybrid} =\mathbf{A}_\text{hybrid}\mathbf{G},
    \label{eq:channel_array_hybrid_unified}
\end{equation} 
where the steering matrix $\mathbf{A}_\text{hybrid}$ is a combination of the \acrlong{ff} and near-field~channels 
\begin{equation}
    \begin{aligned}
    \mathbf{A}_\text{hybrid} &= \sqrt{\frac{M}{L}} \big[\mathbf{a}(\phi_1),\cdots,\mathbf{a}(\phi_{L_o}), \\
    & \qquad \qquad \qquad\mathbf{a}(\phi_{L_o+1},r_{L_o+1}),\cdots,\mathbf{a}(\phi_{L},r_{L}) \big].
    \label{eq:hybrid_polar_domain_transform_matrix}
    \end{aligned}
\end{equation}
where $L_o$ and $L - L_o$ represent the number of far-field paths and near-field paths, respectively. For clarity, we introduce the notation $L_f = L_o$ and $L_n = L - L_o$ in the following sections of the paper. We refer the readers to \cite{Wei2022} for further details on these channel models. 

\section{Proposed Residual Attention CNN Network}\label{sec:propose}
\vspace{-0.5cm}
\begin{figure*}
    \centering
    \begin{subfigure}[b]{0.23\textwidth}
     \centering
     \includegraphics[width=\textwidth]{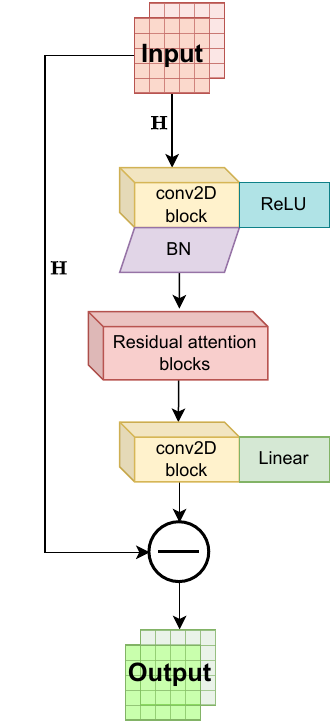}
     \caption{}
     \label{fig:racnn1}
    \end{subfigure}
    \hfill
    \begin{subfigure}[b]{0.23\textwidth}
     \centering
     \includegraphics[width=\textwidth]{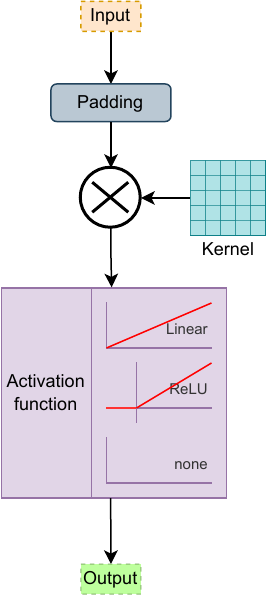}
     \caption{}
     \label{fig:racnn2}
    \end{subfigure}
    \hfill
    \begin{subfigure}[b]{0.23\textwidth}
     \centering 
     \includegraphics[width=\textwidth]{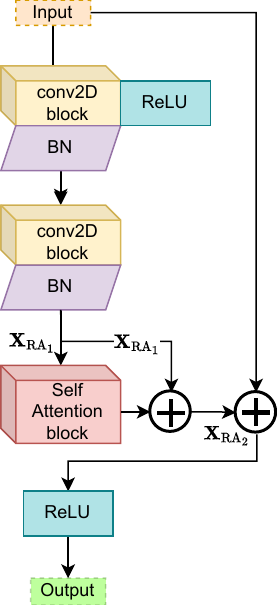}
     \caption{}
     \label{fig:racnn3}
    \end{subfigure}
    \hfill
    \begin{subfigure}[b]{0.23\textwidth}
     \centering
     \includegraphics[width=\textwidth]{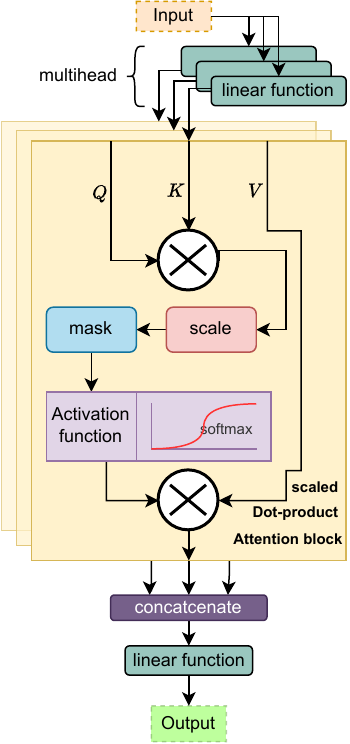}
     \caption{}
     \label{fig:racnn4}
    \end{subfigure}
    \hfill
    \caption{The proposed RACNN network: a) main workflow; b) conv2D block; c) residual attention block; d) self-attention block.}
    \label{fig:racnn}
    \vspace*{-0.5cm}
\end{figure*}

In this section, we propose a new residual attention CNN (RACNN) architecture to estimate the \acrlong{hf} channel in NFC-6G systems. In particular, RACNN consists of four types of blocks: conv2D, residual attention (RA), rectified linear unit (ReLU), and batch normalization (BN). See Fig.~\ref{fig:racnn} for an illustration of our proposed network. 

To simplify the design, we consider a single-user scenario with $K=1$, reducing the channel matrix $\mathbf{H}$ to a vector representation $\mathbf{h} \in \mathbb{C}^{M \times 1}$. Initially, RACNN transforms this vector into a matrix representation, expressed as $\mathbf{h} \in \mathbb{C}^{M \times1} \rightarrow \mathbf{X} \in \mathbb{R}^{2 \times N_x \times N_y}$, where $N_x \times N_y = M$. This transformation separates the complex elements of $\mathbf{h}$ into their real and imaginary components, which facilitates more effective training~\cite{gao2024}. At this stage, $\mathbf{X}$ is represented as a 2-channel image and serves as the input for the channel estimation process. As illustrated in Fig.~\ref{fig:racnn}(a) and Fig.~\ref{fig:racnn}(b), RACNN incorporates multiple 2D convolutional (conv2D) blocks which are mathematically expressed as  
\begin{equation}
    \mathbf{Y}(C_{y})=  \sum_{k=0}^{C_{x}-1}\mathbf{W}(C_{y},k) \star \mathbf{X}(k) + \texttt{bias}(C_{y}).
    \label{eq:CNN_formular}
\end{equation} 
Here, $``\star"$ denotes the 2D convolution operator, $\mathbf{W}(C_y, k) \in \mathbb{R}^{r \times r}$ (e.g., $r=3$) is the convolution kernel connecting $k$-th input channel to $C_y$-th output channel, and $\texttt{bias}(C_y) \in \mathbb{R}$ is the associated bias term. 
In our implementation, the first convolutional layer takes $C_x = 2$ as input, and the number of output channels is set to $C_y = 64$. In subsequent layers, the number of input and output channels remains fixed at $C_x = C_y = 64$. At the final layer, the output channels are reduced to $C_y = 2$, yielding an estimated noise that shares the same shape as the input $\mathbf{X} \in \mathbb{R}^{2 \times N_x \times N_y}$.

To reduce overfitting during the training process, the output from certain conv2D blocks passes through a ReLU activation function~\cite{Ide2017ReLU} to introduce nonlinearity. BN~\cite{Ioffe2015} is also employed to adjust and scale activations, helping reduce the covariate shift. Specifically, the BN in our network is defined by
\begin{equation}
    {\mathbf{Y}_\text{BN}} = \gamma \hat{\mathbf{Y}}_\text{conv2d} + \beta,
    \label{eq:mini_batch_scalenshift}
\end{equation}
where $\gamma$ and $\beta$ are the scale and shift parameters, ${\hat{\mathbf{Y}}_\text{conv2d}}$ is normalized from the input ${\mathbf{Y}_\text{conv2d}}$ (i.e., output of the conv2d layer) as  ${\hat{\mathbf{Y}}_\text{conv2d}} = \frac{{\mathbf{Y}_\text{conv2d}}-\mu_B}{\sqrt{\sigma^2_B+\epsilon}}$, with $\mu_B$ and $\sigma^2_B$ is the mean and variance of ${\mathbf{Y}_\text{conv2d}}$.

To support the extraction stage in each conv2D, an RA block is paired with the network to introduce the attention mechanism. The \acrshort{ra} block has two mechanisms: (i) the data attention function refines the features learned by the network, and (ii) the residual connection function helps maintain effective network learning as data progresses deeper~\cite{Wang2017}. Particularly in our RACNN network, we use the \textbf{self-attention mechanism} based on~\cite{Vaswani2017}. This layer generates an attention map that highlights the important regions or features of the input using its own weights. Accordingly, \acrshort{ra} block helps each output from conv2D blocks retain the original information while incorporating the refined features from the attention mechanism. 
Fig.~\ref{fig:racnn}(c) illustrates the architecture of an RA block, which follows the following sequence:

(i) The attention layer first calculates the Query, Key, and Value by applying the learned linear transformation to an input matrix $\mathbf{X}_{\text{RA}_1}$, as illustrated in Fig.~\ref{fig:racnn}(d). As self-attention takes the same weights from all inputs, this remains consistent across all heads
\begin{equation}
    \mathbf{Q}=\mathbf{X}_{\text{RA}_1}\mathbf{W}_\mathbf{Q},\quad\mathbf{K}=\mathbf{X}_{\text{RA}_1}\mathbf{W}_\mathbf{K}, \quad\text{and} \quad\mathbf{V}=\mathbf{X}_{\text{RA}_1}\mathbf{W}_\mathbf{V}.
    \label{eq:self_attention_weight_calculation}
\end{equation}
Here, $\left\{\mathbf{Q,\,K,\,V}\right\}$ represent the query, key, and value matrices, respectively, and $\mathbf{W}_{\mathbf{Q}},  \mathbf{W}_{\mathbf{K}}, \mathbf{W}_{\mathbf{V}}$ denote their attention weight matrices.  The attention weights are computed by comparing the query and key vectors. The difference between these two vectors determines how much attention an input element should give to others. This is typically done by calculating the dot product between the query and key vectors:
\begin{equation}
    \mathbf{W}_\text{attention} = \text{softmax}\left( \frac{\mathbf{Q}\mathbf{K}^\top}{\sqrt{D}}\right),
\end{equation}
where the scaling factor $\sqrt{D}$ is introduced to prevent large values that may cause instability. Then, the output $\mathbf{X_\text{attention}}$ of the attention mechanism is obtained~by
\begin{equation}
    \mathbf{X_\text{attention}}= \mathbf{W}_\text{attention} \times \textbf{V},
\end{equation}

(ii) The attention details are then combined with the original input to perform a residual connection operation. Similar to the ResNet~\cite{He2016ResNet} model, the original input $\mathbf{X}_{\text{RA}_1}$ is added to the attention feature map $\mathbf{X_\text{attention}}$, resulting in the final output:
\begin{equation}
    \mathbf{X}_{\text{RA}_2} = \mathbf{X}_{\text{RA}_1} + \mathbf{X_\text{attention}}.
\end{equation}

\section{Numerical Experiments}\label{sec:result}

In this section, we conduct various experiments with different channel configurations to evaluate the performance of RACNN. We also compare RACNN with traditional methods, including LS and MMSE~\cite{Kay1993}, as well as the deep learning method \acrshort{xlcnet}~\cite{gao2024}. Several channel configurations will be used for validation, including far-field, near-field, and hybrid channels.

\subsection{Experiment Setup}

For training, both \acrshort{racnn} and \acrshort{xlcnet} will use the same channel configuration dataset, similar to work in \cite{gao2024}. Particularly, the experimental parameters are detailed in Table~\ref{tab:params}. All experiments are conducted on a workstation computer with an Intel Core i9-14900K processor, 128GB of RAM, and an Nvidia RTX~4090.

\begin{table}[!t]
\centering
\caption{Simulation parameters of the training process and wireless communications system.}
\vspace{1em}
\label{tab:params}
\begin{tblr}{
  width = \linewidth,
  colspec = {Q[300]Q[200]Q[300]Q[200]},
  row{1} = {c},
  hlines,
  vlines,
}
\textbf{Parameters} & \textbf{\textbf{\textbf{\textbf{Specification}}}} & \textbf{\textbf{Parameters}} & \textbf{\textbf{Specification}}\\
Signal frequency & $f=30$GHz & No. training samples & 180000~ ~ ~\\
Antenna size & $N=256$ & No. testing samples & 60000\\
Wavelength & $\lambda=0.01$ & No. iterations & 200\\
Antenna distance & $\lambda / 2$ & Learning rate & 0.001\\
No. far-field paths & $L_f = \mathcal{U}\left[0;10\right]$ & Optimizer & Adam\\
No. near-field paths & $L_n = \mathcal{U}\left[0;10\right]$ & Loss function & MSE\\
Near-field distance & $r_l = \mathcal{U}(10; 80)$ & Training SNR levels & $10$ and $15$dB
\end{tblr}
\vspace{-0.5cm}
\end{table}

To evaluate the algorithm's performance, we use the \acrfull{nmse} metric to validate the performance over SNR levels, and  the Mean Square Error (MSE) to calculate loss over training iterations: 
\begin{equation}
    \text{NMSE} = \frac{\sum_{i=1}^{n} (y_i - \hat{y}_i)^2}{\sum_{i=1}^{n} |y_i|^2}, \qquad \text{MSE} = \frac{1}{n}\sum_{i=1}^n (y_i - \hat{y_i})^2,
    \label{eq:normalize&based_mean_square_error}
\end{equation}
where $y_i$ and $\hat{y}_i$ denote the ground truth and estimated values, respectively, and $n$ is the total number of observations.

\subsection{Experimental Results}
\subsubsection{Evaluation on Synthetic Datasets:}
\label{ex:1}

\begin{figure}[!t]
    \centering
     \begin{subfigure}[t]{0.45\textwidth}
        \centering
        \includegraphics[width=\linewidth]{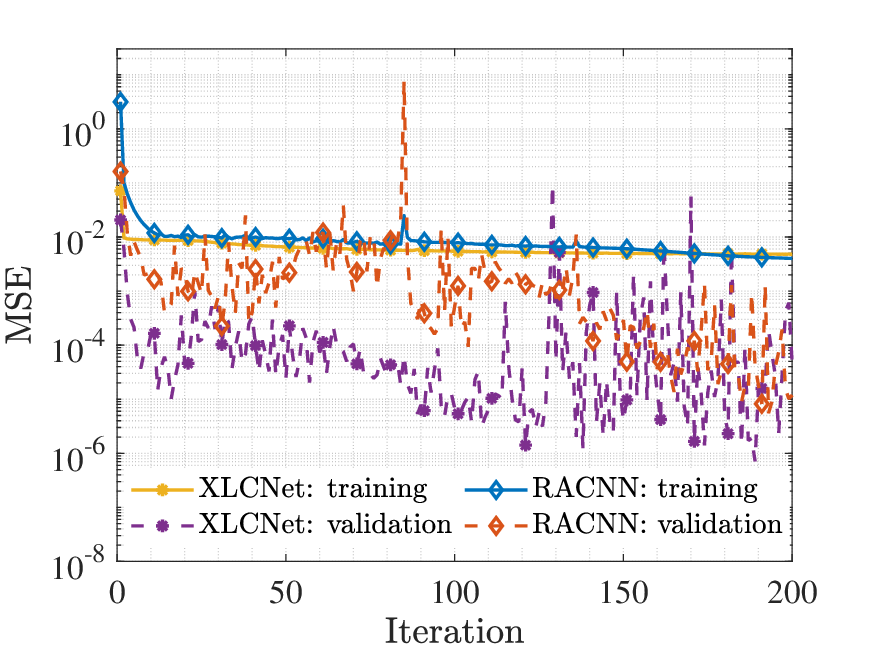}
        \caption{Training and validation losses}
    \end{subfigure}
    \begin{subfigure}[t]{0.45\textwidth}
        \centering
        \includegraphics[width=\linewidth]{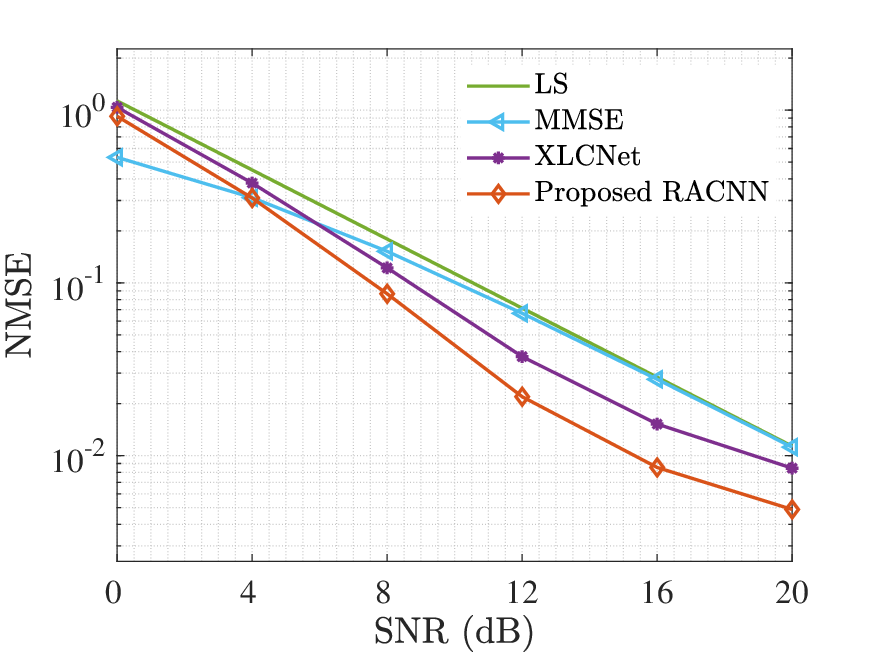}
        \caption{Performance on testing dataset}
    \end{subfigure}
    \caption{Performance of training and testing phase.}
    \label{fig:performance_testing_using_training_dataset}
    \vspace*{-0.5cm}
\end{figure}

Fig.~\ref{fig:performance_testing_using_training_dataset} shows the testing results using the synthetic dataset from Table~\ref{tab:params}. We can see that  RACNN outperforms other methods in most scenarios. Specifically, RACNN demonstrates improved estimation accuracy performance compared to the deep learning model XLCNet (a CNN model) across all scenarios. This difference is particularly noticeable at higher SNRs, ranging from $16$dB to $20$dB, where the error rate gap can be as large as $0.5$.  However, when considering data extraction speed and training rate, XLCNet has the advantage. As illustrated in Fig.~\ref{fig:performance_testing_using_training_dataset}(a), RACNN has a slightly slower loss convergence speed compared to XLCNet (as shown by the training loss curve). In terms of validation,  the loss during testing on RACNN remains higher for more iterations and only reaches optimal results around $150$~iterations. This suggests an increased computational requirement due to the integration of \acrshort{ra} blocks.

\subsubsection{Far-Field and Near-Field Channel Estimation:} 
\label{ex:2}

\begin{figure}[!t]
    \centering
    \begin{subfigure}[t]{0.30\textwidth}
        \centering
        \includegraphics[width=\textwidth]{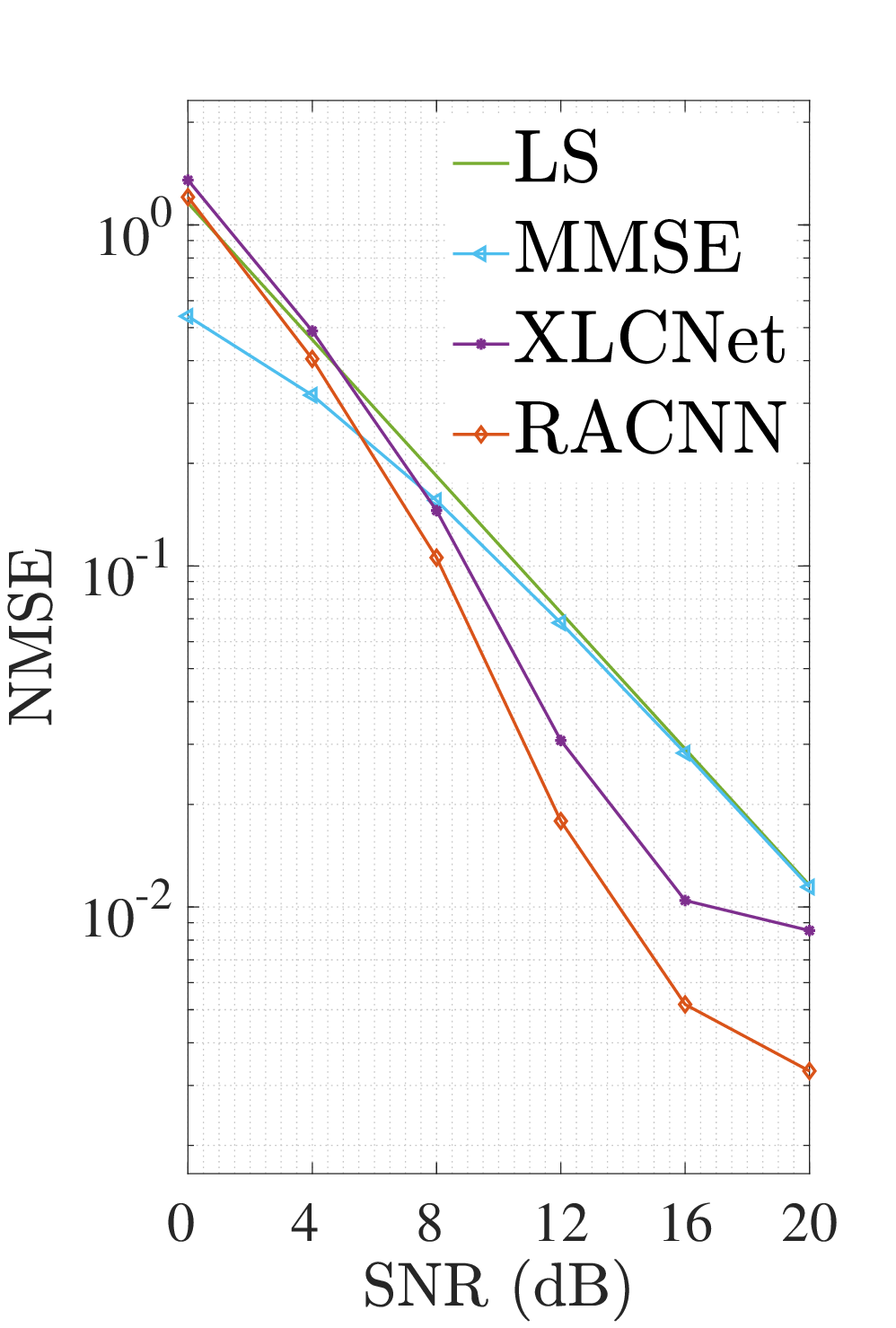}
        \caption{$L_f = 3$}
    \end{subfigure}
    \begin{subfigure}[t]{0.30\textwidth}
        \centering
        \includegraphics[width=\textwidth]{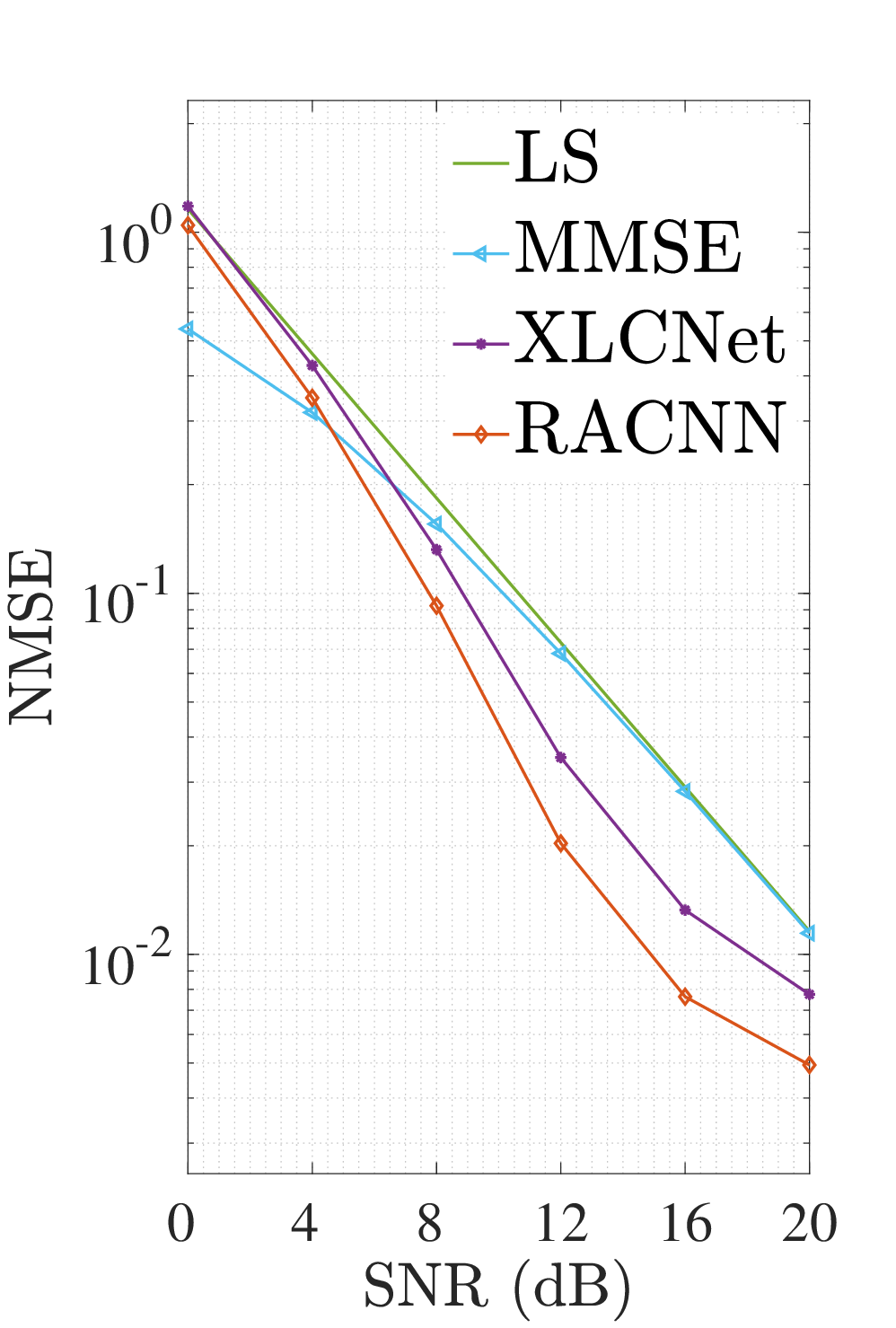}
        \caption{$L_f = 5$}
    \end{subfigure}
    \begin{subfigure}[t]{0.30\textwidth}
        \centering
        \includegraphics[width=\textwidth]{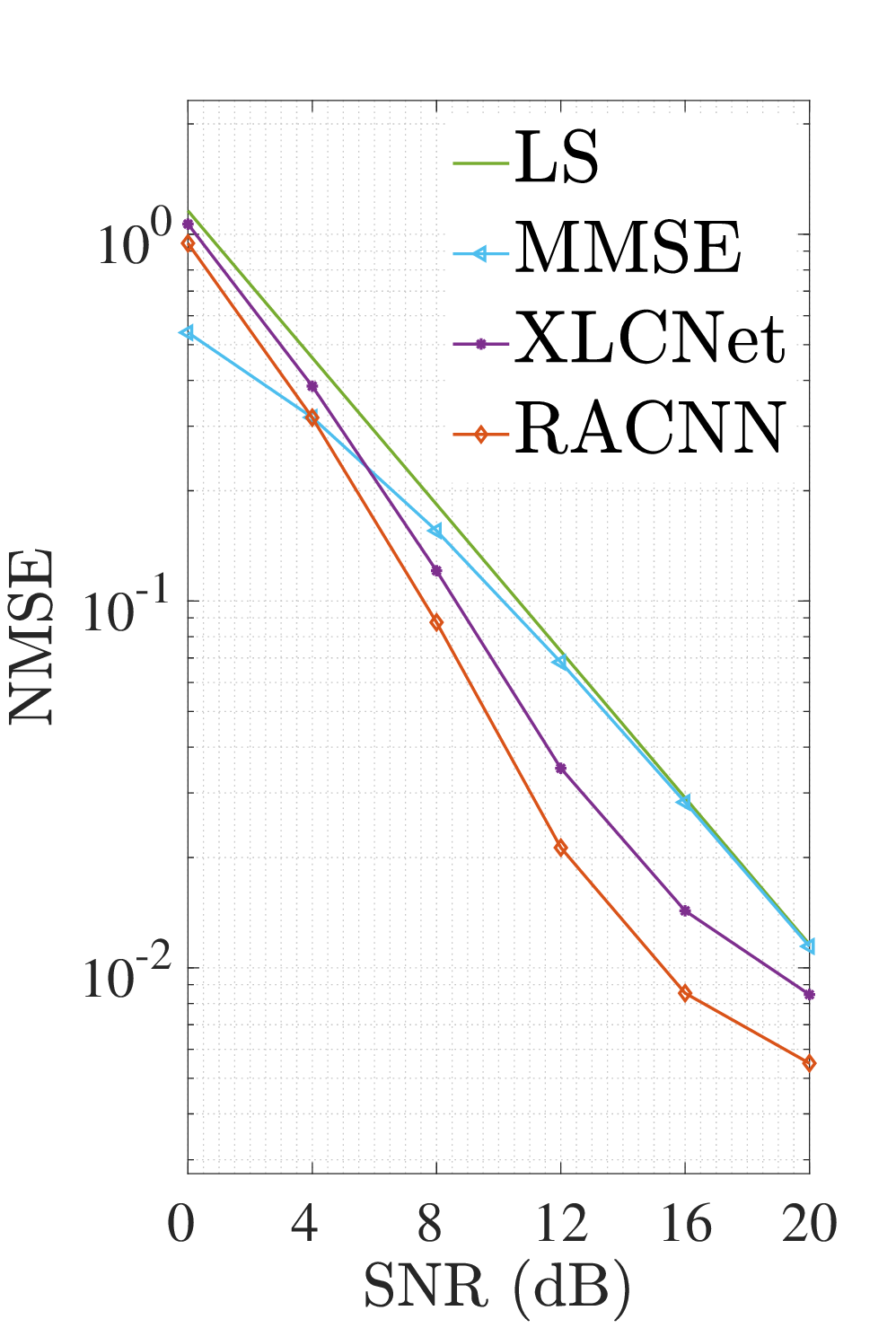}
        \caption{$L_f = 7$}
    \end{subfigure}        
    \caption{Performance of estimators for far-field only.}
    \label{fig:Comparing_different_method_far-field_only}
    \vspace*{-0.5cm}
\end{figure}

\begin{figure}[!t]
    \centering
    \begin{subfigure}[t]{0.30\textwidth}
        \centering
        \includegraphics[width=\textwidth]{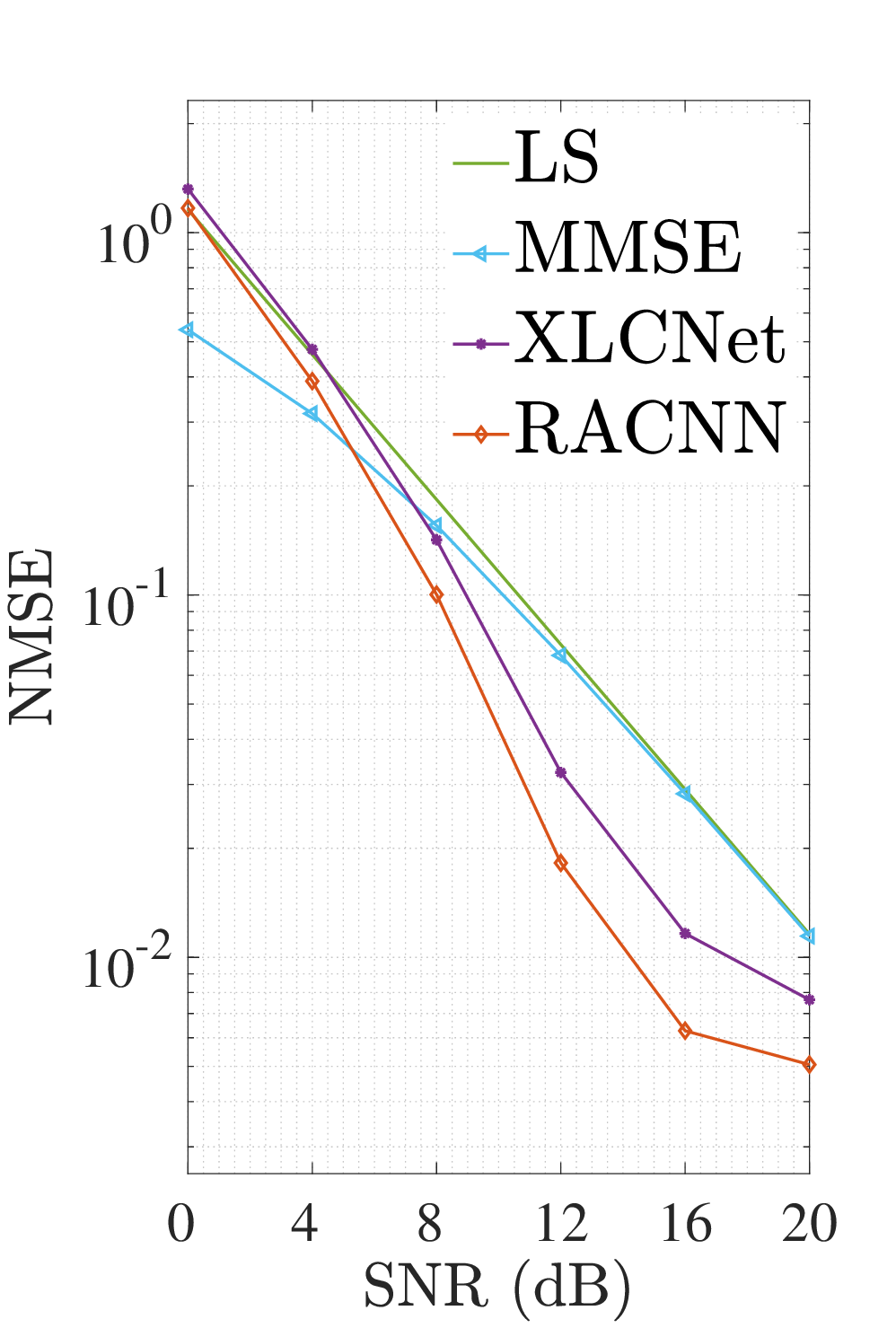}
        \caption{$L_n = 3$}
    \end{subfigure}
    \begin{subfigure}[t]{0.30\textwidth}
        \centering
        \includegraphics[width=\textwidth]{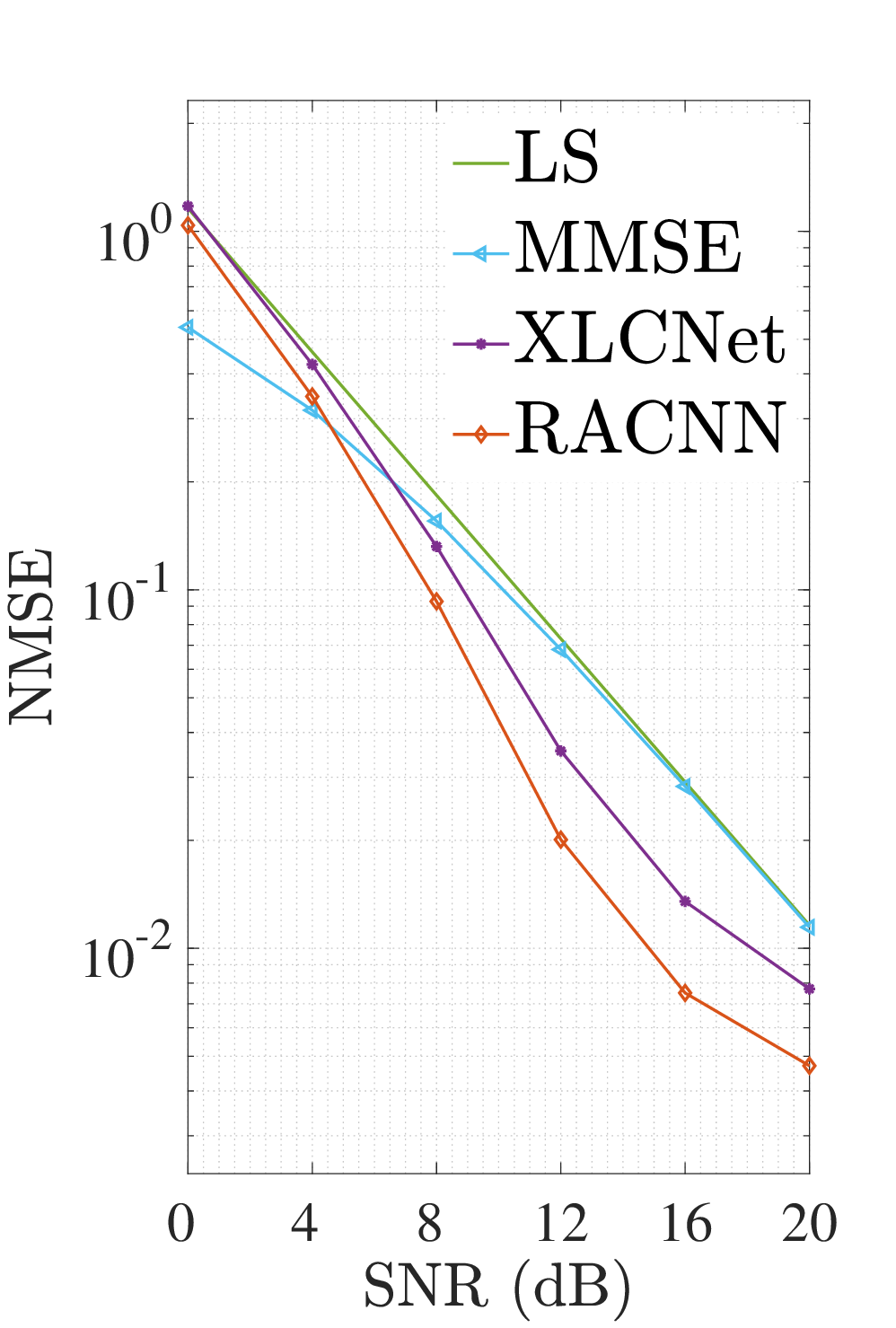}
        \caption{$L_n = 5$}
    \end{subfigure}
    \begin{subfigure}[t]{0.30\textwidth}
        \centering
        \includegraphics[width=\textwidth]{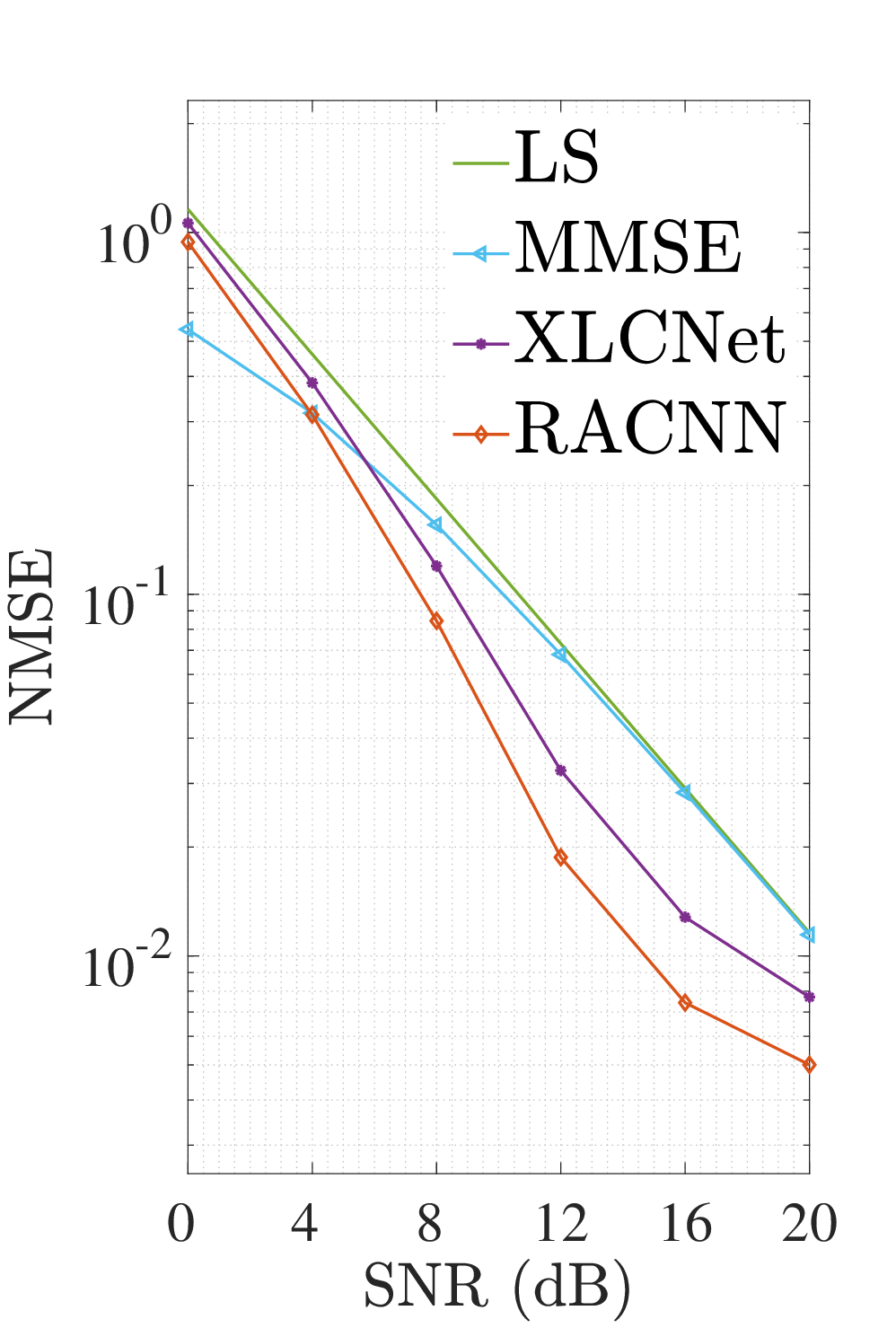}
        \caption{$L_n = 7$}
    \end{subfigure}   
    \caption{Performance of estimators for near-field only.}
    \label{fig:Comparing_different_method_near-field_only}
    \vspace*{-0.5cm}
\end{figure}
\begin{figure}[!t]
    \centering
    \begin{subfigure}[t]{0.45\textwidth}
        \centering
        \includegraphics[width=\textwidth]{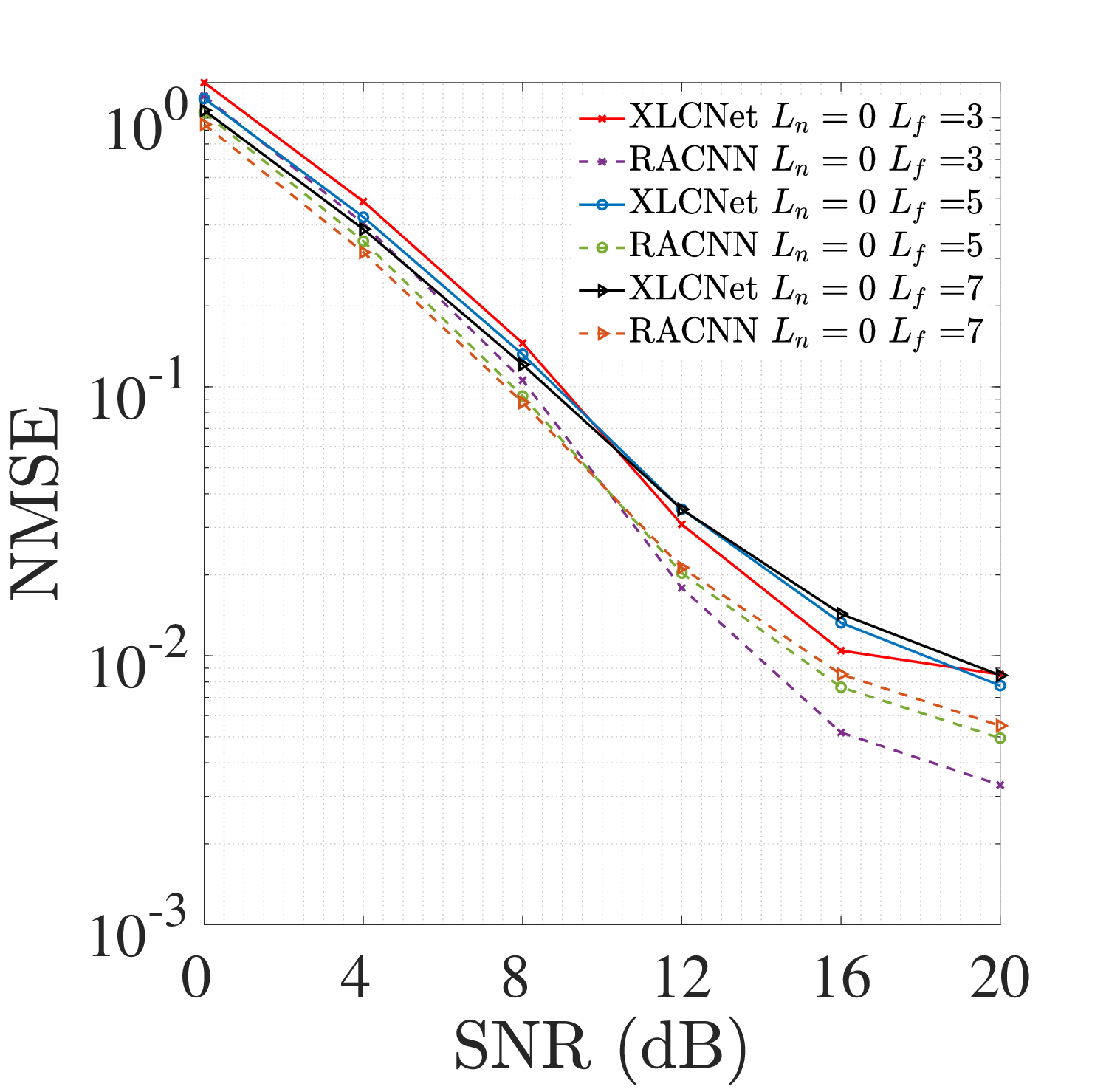}
        \caption{Far-field scenarios}
    \end{subfigure}  
    \begin{subfigure}[t]{0.45\textwidth}
        \centering
        \includegraphics[width=\textwidth]{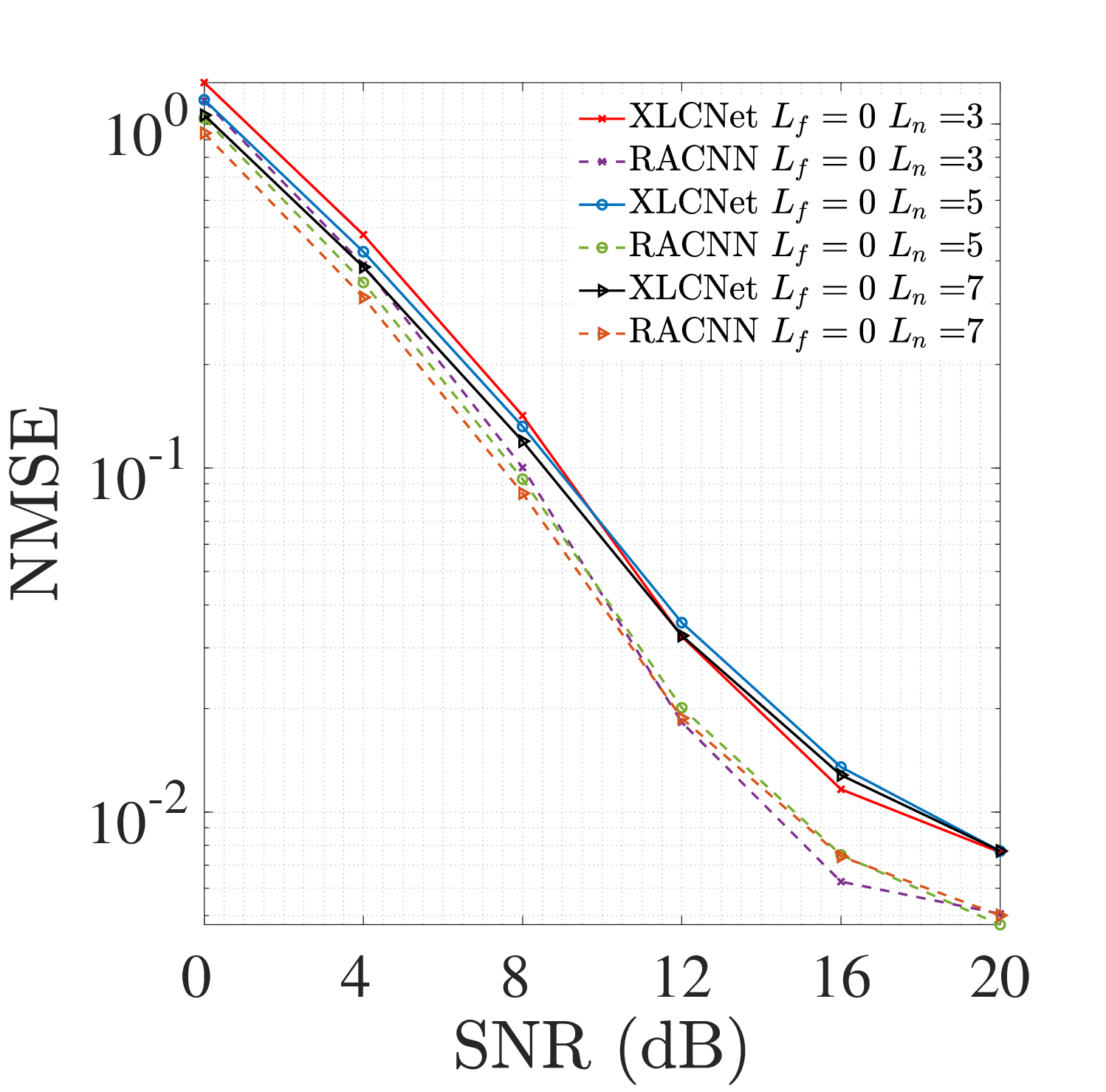}
        \caption{Near-field scenarios}
    \end{subfigure}    
    \caption{Performance of estimators in scenerio for far-/\acrlong{nf} only.}
    \label{fig:Detail_compare_single_channel}
    \vspace*{-0.5cm}
\end{figure}

In this experiment, we validate the performance of RACNN on \acrlong{nf} channel and \acrlong{ff} channel independently. This is done by modifying the BS transmit path $L_f$ and $L_n$ of each electromagnetic field in a range of constant (3, 5, and~7). The experimental results are shown in Fig.~\ref{fig:Comparing_different_method_far-field_only} and Fig.~\ref{fig:Comparing_different_method_near-field_only}. They indicate that RACNN achieves the best estimation accuracy performance in most testing cases. At low SNRs ($\leq$4dB), MMSE performs better, particularly with lower multipath values $L_f = 3$ and $L_n = 3$. However, when the number of multipaths increases, the performance of RACNN begins to improve significantly, as compared to others. Fig.~\ref{fig:Detail_compare_single_channel} shows the detailed results, investigating channel configurations. We can observe that all methods perform poorly at lower SNR levels. As the SNR increases, a noticeable difference in performance emerges. Looking at the \acrlong{ff} channel, changes in channel configuration have a small impact on XLCNet's results. In contrast, with RACNN, each line becomes progressively more differentiated as the SNR increases. This behavior differs when the same test is conducted for the \acrlong{nf} channel, where both RACNN and XLCNet exhibit stable performance across all scenarios despite the increasing SNR.

\subsubsection{Hybrid-Field Channel Estimation:} 

\label{ex:3}
\begin{figure}[!t]
    \centering
    \begin{subfigure}
        [t]{0.45\textwidth}
        \centering
        \includegraphics[width=\textwidth]{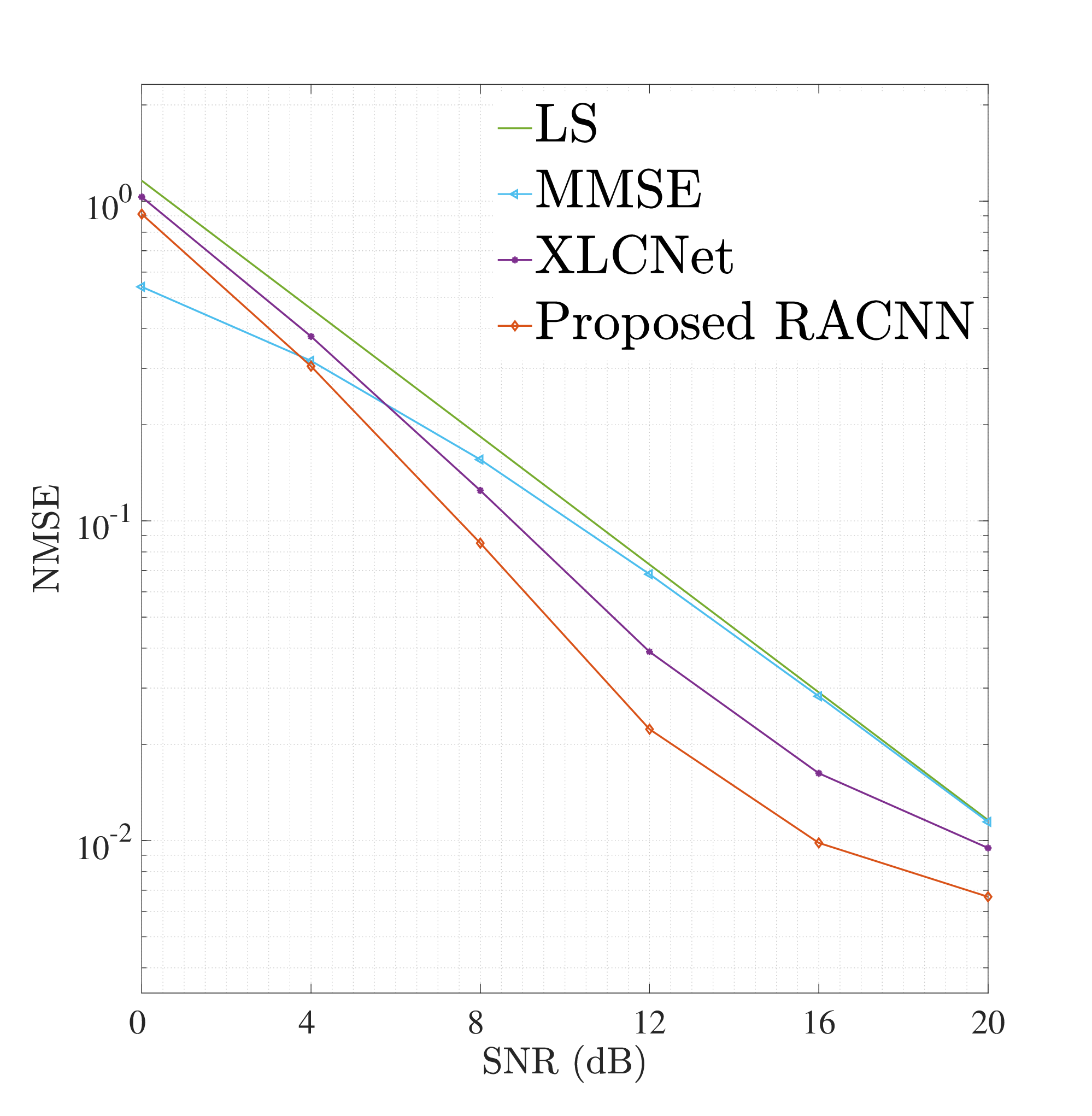}
        \caption{$L_f = 6; L_n = 3$}
    \end{subfigure}
    \begin{subfigure}
        [t]{0.45\textwidth}
        \centering
        \includegraphics[width=\textwidth]{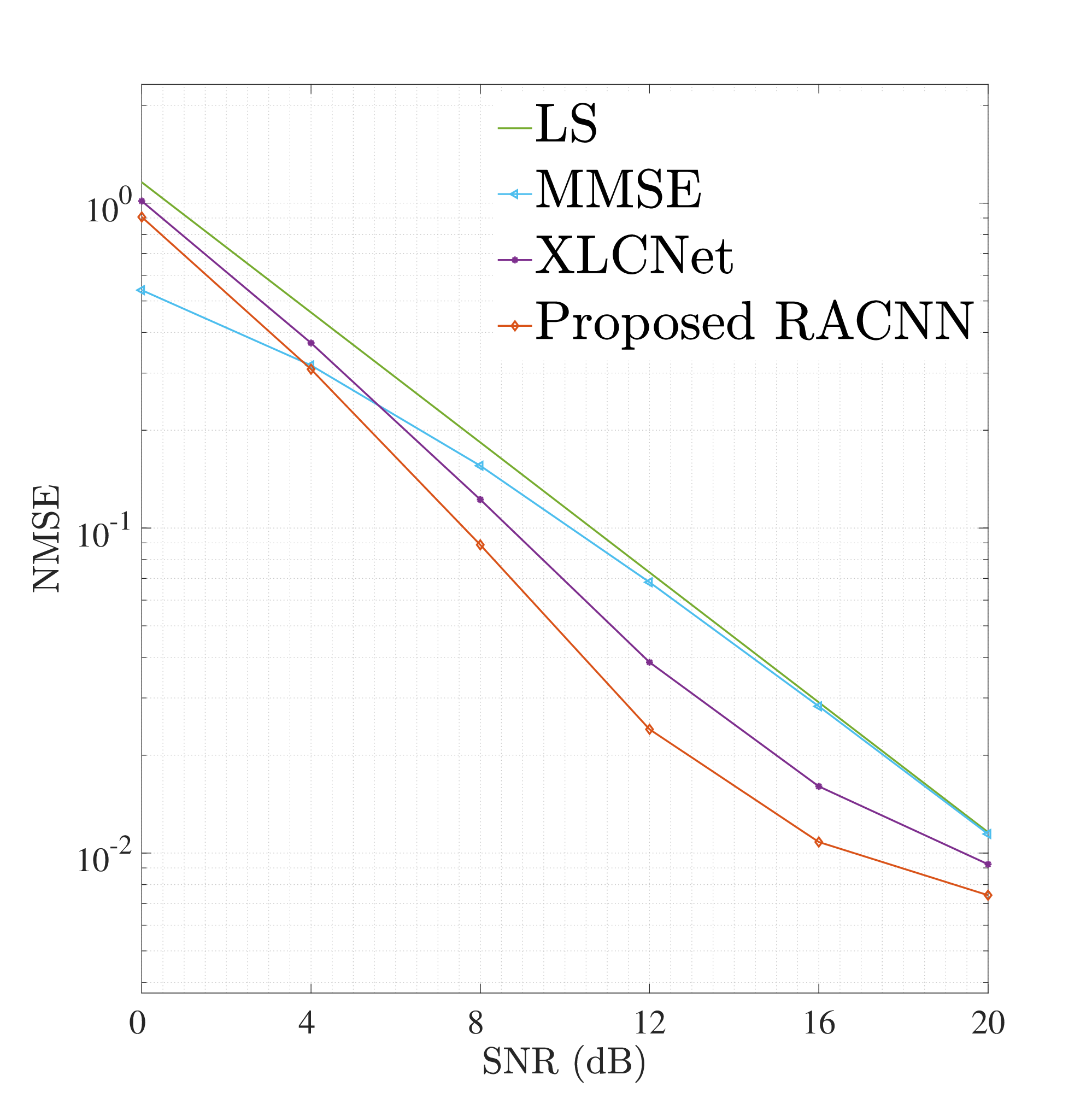}
        \caption{$L_f = 9; L_n = 3$}
    \end{subfigure}
    \caption{Performance of hybrid channel model with varying in \acrlong{ff}.}
    \label{fig:Comparing_different_method_hybrid_far-field_increase}
    \vspace*{-0.5cm}
\end{figure}
\begin{figure}[!t]
    \centering
    \begin{subfigure}
        [t]{0.45\textwidth}
        \centering
        \includegraphics[width=\textwidth]{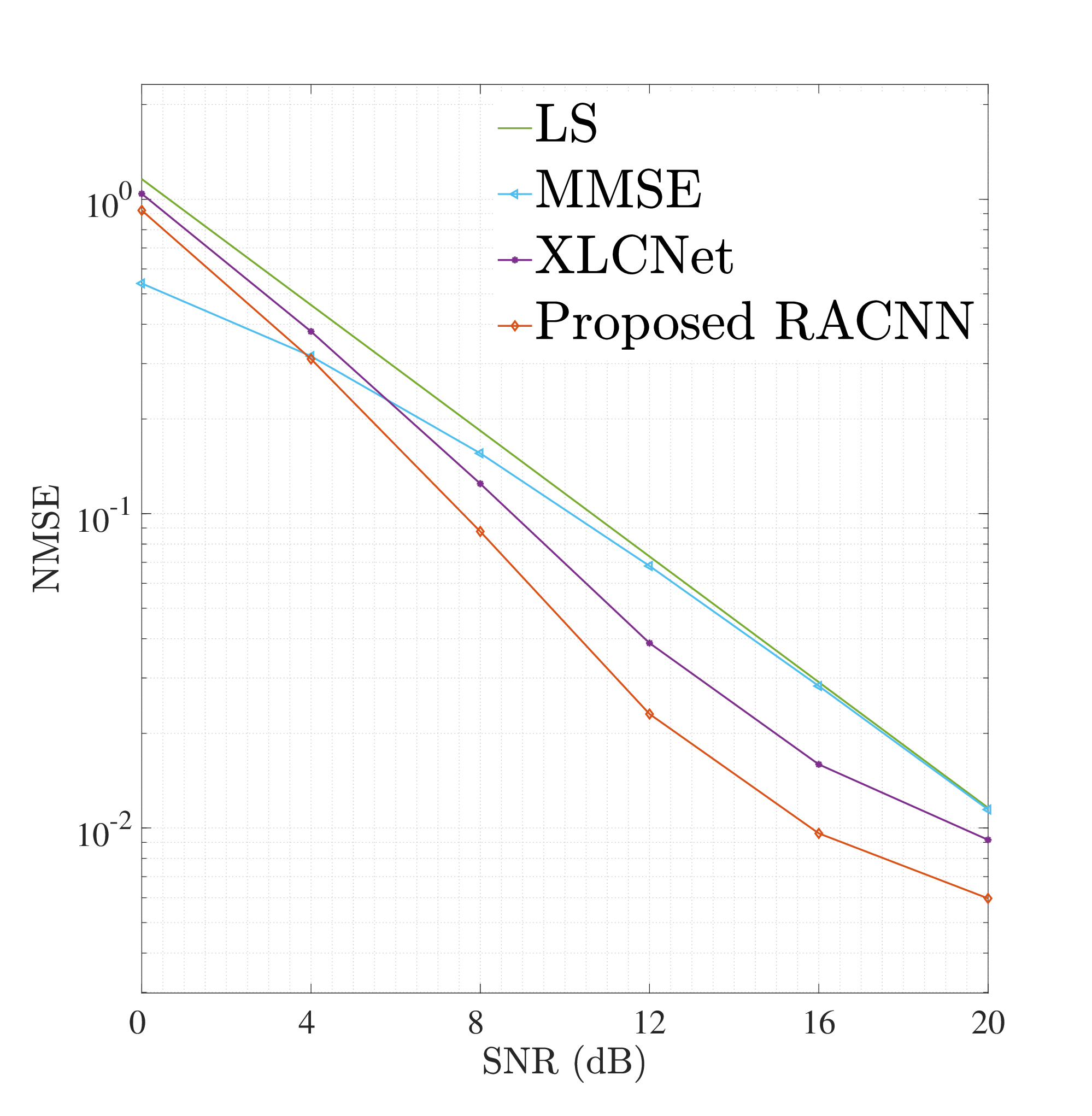}
        \caption{$L_f = 3; L_n = 6$}
    \end{subfigure}
    \begin{subfigure}
        [t]{0.45\textwidth}
        \centering
        \includegraphics[width=\textwidth]{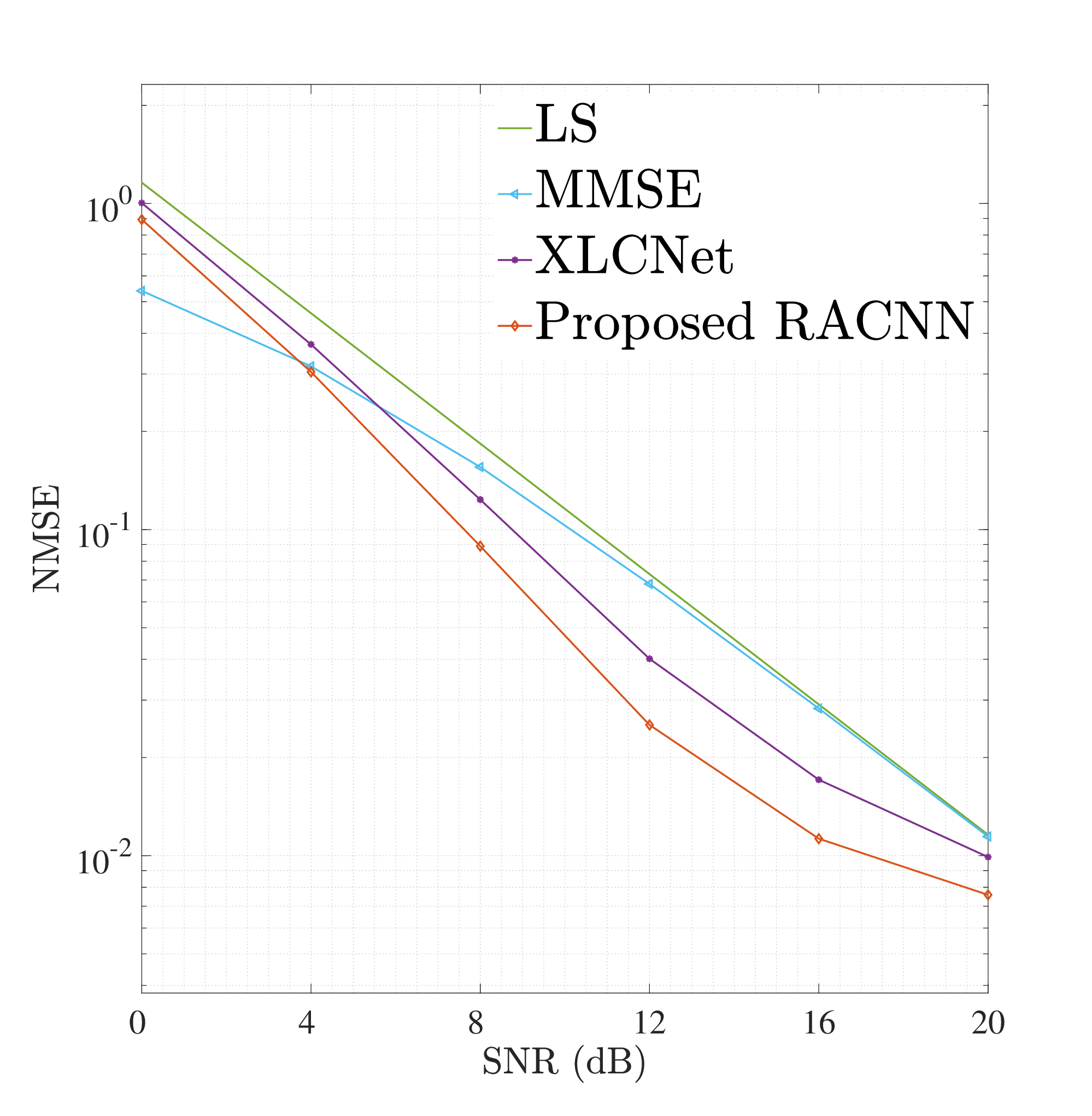}
        \caption{$L_f = 3; L_n = 9$}
    \end{subfigure}
    \caption{Performance of hybrid channel model with varying in \acrlong{nf}.}
    \label{fig:Comparing_different_method_hybrid_near-field_increase}
    \vspace*{-0.5cm}
\end{figure}

\begin{figure}[!t]
    \centering
    \begin{subfigure}[t]{0.45\textwidth}
        \centering
        \includegraphics[width=\textwidth]{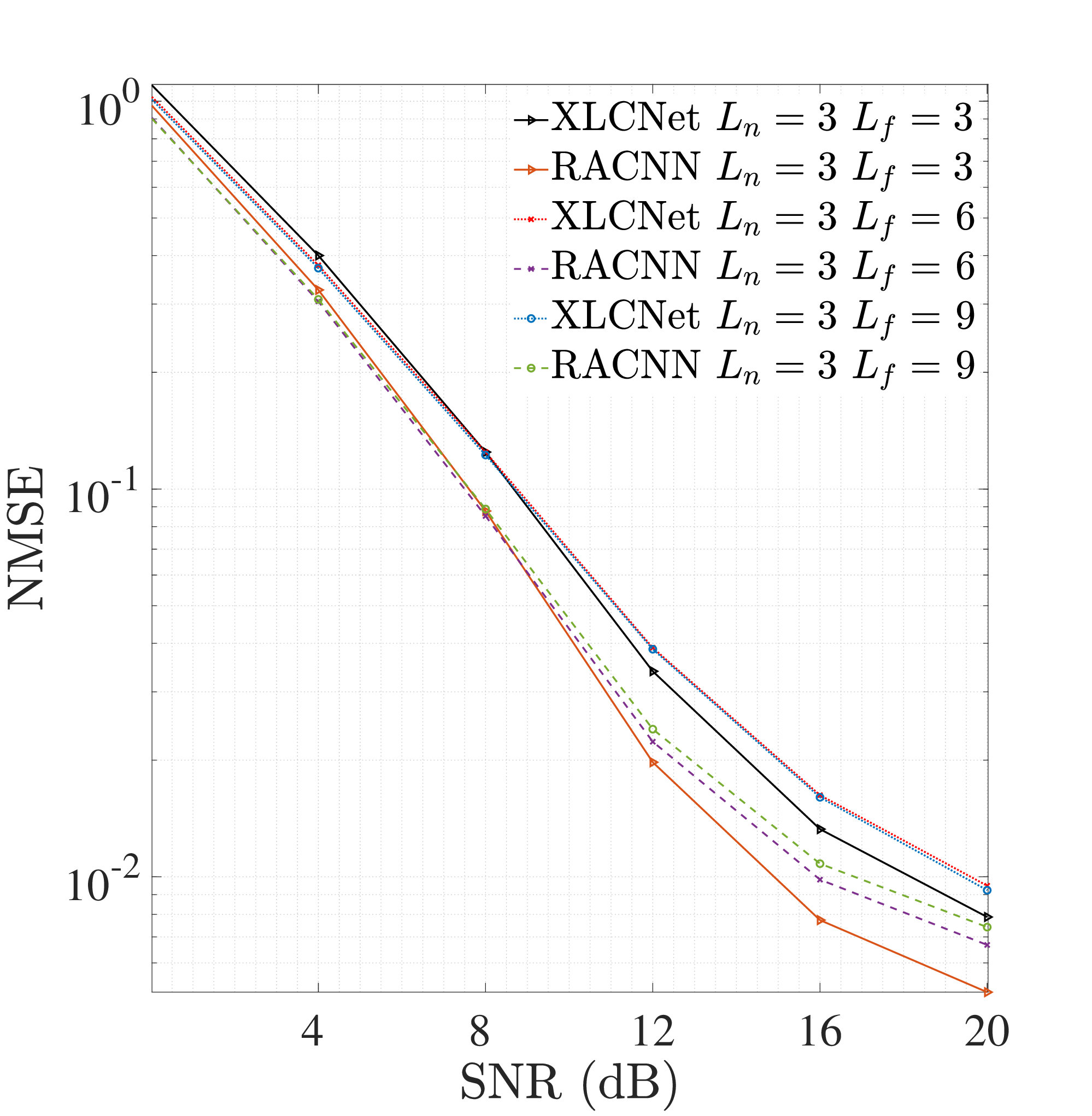}
        \caption{Different number of \acrlong{ff} paths}
    \end{subfigure}  
    \begin{subfigure}[t]{0.45\textwidth}
        \centering
        \includegraphics[width=\textwidth]{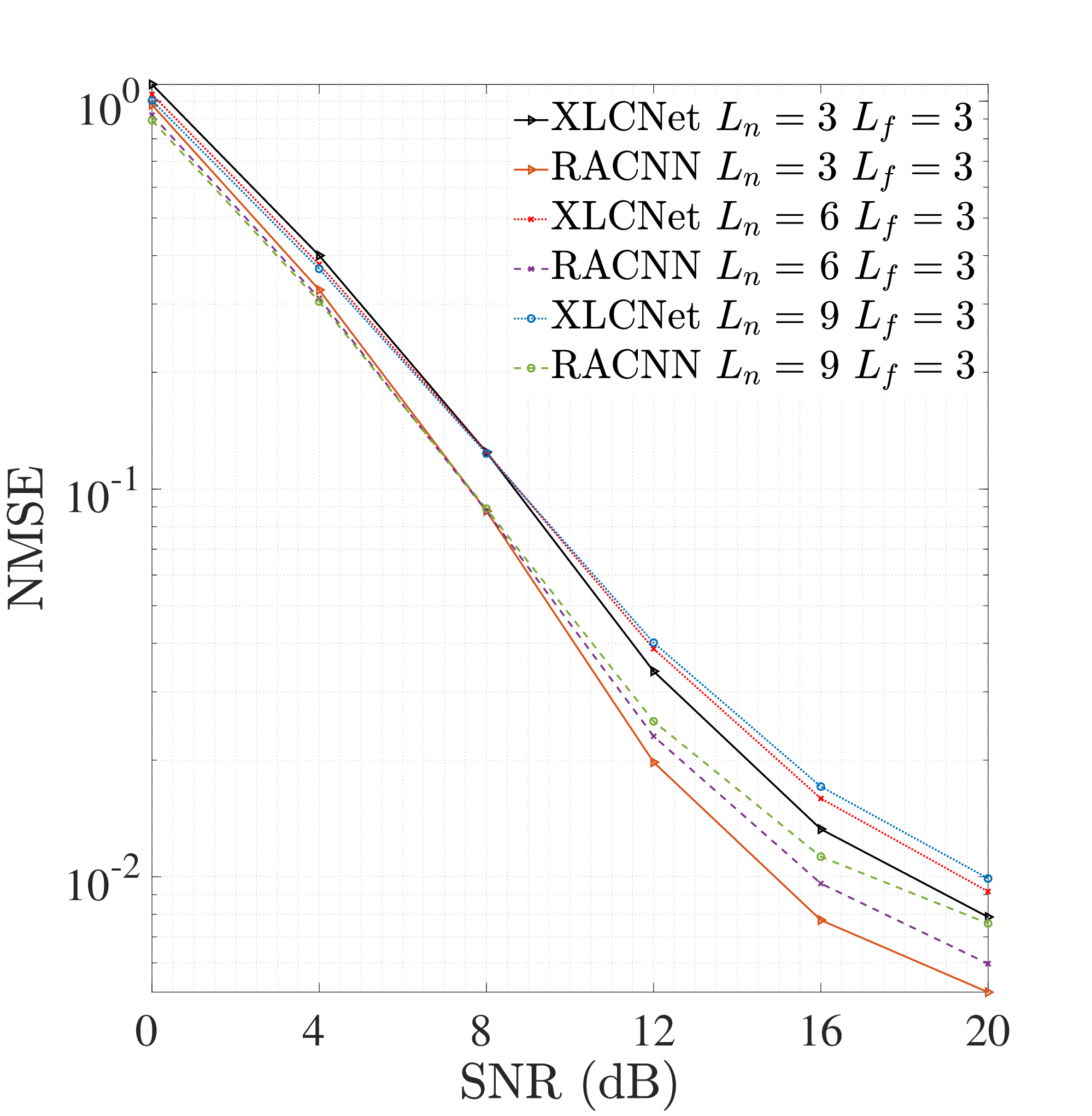}
        \caption{Different number of \acrlong{nf} paths}
    \end{subfigure}    
    \caption{Performance of RACNN and XLCNet in estimating hybrid channels.}
    \label{fig:Detail_compare_hybrid_channel}
    \vspace*{-0.5cm}
\end{figure}

Similar to the experiments above, we also apply the same validation approach for the \acrlong{hf} channel by mixing different $L_f$ and $L_n$ settings. The test begins with the lowest multipath configurations at $L_f=3$ and $L_n=3$, then gradually increases far-field and near-field paths to different values (e.g., 6 or 9) to investigate the impact of system configuration changes on RACNN and its competitors. The experimental results, shown in Fig.~\ref{fig:Detail_compare_hybrid_channel}, indicate that RACNN maintains its position of the best performance on \acrlong{hf} channel except at SNR of 0dB. Consistent with the results of previous experiments, RACNN's performance improves as SNR increases, outperforming XLCNet at high SNRs. From Fig.~\ref{fig:Detail_compare_hybrid_channel}, we also observe that RACNN is a good option for \acrlong{hf} channel estimation, achieving good performance, especially at high SNRs with fewer multipaths. The difference in RACNN performance is expressed prominently at the high SNR range, especially in the range of 16 to 20dB, where the gap between the best and worst results reaches approximately $2*10^{-3}$. From here, we expect a potential where RACNN capability could be improved from the current result with a few tweaks in model architecture. Despite the performance differences, an increase in channel paths does not directly lead to noise reduction, as performance ranking remains inconsistent across SNR levels. This is evident in Fig.~\ref{fig:Detail_compare_hybrid_channel}, where we observed the advantage of increased transmitting multipaths at lower SNR. However, as SNR rises, channel configurations with higher multipaths begin to pose a drawback. This property applies to both far-field and near-field channels.

\section{Conclusion}\label{sec:concl}

In this paper, we proposed the Residual Attention Convolutional Neural Network for near-field channel estimation in 6G wireless communication systems. RACNN integrates convolutional layers with self-attention mechanisms to enhance feature extraction during denoising process. Through various experiments, RACNN consistently outperformed other estimators, i.e., LS, MMSE, and XLCNet. More details, in scenarios with different far-field and near-field path settings, RACNN maintained the best performance across all cases, except at 0dB SNR where MMSE had a slight advantage. The experimental data also indicated that RACNN's performance improved significantly as the number of near-field and far-field paths reduced. Overall, the findings suggest that RACNN is a promising solution for the complex signal processing demands of ELAA in 6G systems, offering robustness and improved accuracy in near-field channel estimation.

\subsubsection{Acknowledgments} 
This research was funded by the research project QG (QG.25.08) of Vietnam National University Hanoi. Correspondence: Tran Thi Thuy Quynh (quynhttt@vnu.edu.vn)

\subsubsection{Disclosure of Interests}
The authors have no competing interests.

%
%
\bibliographystyle{./styles/splncs04}
\bibliography{library}

\begin{thebibliography}{10}
\providecommand{\url}[1]{\texttt{#1}}
\providecommand{\urlprefix}{URL }
\providecommand{\doi}[1]{https://doi.org/#1}

\bibitem{cui2022near}
Cui, M., Wu, Z., Lu, Y., Wei, X., Dai, L.: Near-field mimo communications for 6g: Fundamentals, challenges, potentials, and future directions. IEEE Communications Magazine  \textbf{61}(1),  40--46 (2023)

\bibitem{gao2024}
Gao, S., Dong, P., Pan, Z., You, X.: Lightweight deep learning based channel estimation for extremely large-scale massive mimo systems. IEEE Transactions on Vehicular Technology  \textbf{73}(7),  10750--10754 (2024)

\bibitem{Hassan2020}
Hassan, K., Masarra, M., Zwingelstein, M., Dayoub, I.: Channel estimation techniques for millimeter-wave communication systems: Achievements and challenges. IEEE Open Journal of the Communications Society  \textbf{1},  1336--1363 (2020)

\bibitem{He2016ResNet}
He, K., Zhang, X., Ren, S., Sun, J.: Deep residual learning for image recognition. In: IEEE Conference on Computer Vision and Pattern Recognition (CVPR). pp. 770--778 (2016)

\bibitem{Ide2017ReLU}
Ide, H., Kurita, T.: Improvement of learning for cnn with relu activation by sparse regularization. In: International Joint Conference on Neural Networks (IJCNN). pp. 2684--2691 (2017)

\bibitem{Ioffe2015}
Ioffe, S., Szegedy, C.: Batch normalization: accelerating deep network training by reducing internal covariate shift. In: 32nd International Conference on International Conference on Machine Learning (ICML). p. 448–456 (2015)

\bibitem{Jiang2021}
Jiang, W., Han, B., Habibi, M.A., Schotten, H.D.: The road towards 6g: A comprehensive survey. IEEE Open Journal of the Communications Society  \textbf{2},  334--366 (2021)

\bibitem{Kay1993}
Kay, S.M.: Fundamentals of statistical signal processing: estimation theory. Prentice-Hall, Inc., USA (1993)

\bibitem{qiang2020}
Qiang, Y., Shao, X., Chen, X.: A model-driven deep learning algorithm for joint activity detection and channel estimation. IEEE Communications Letters  \textbf{24}(11),  2508--2512 (2020)

\bibitem{Sun2025}
Sun, S., Li, R., Han, C., Liu, X., Xue, L., Tao, M.: How to differentiate between near field and far field: Revisiting the rayleigh distance. IEEE Communications Magazine  \textbf{63}(1),  22--28 (2025)

\bibitem{Vaswani2017}
Vaswani, A., Shazeer, N., Parmar, N., Uszkoreit, J., Jones, L., Gomez, A.N., Kaiser, {\L}., Polosukhin, I.: Attention is all you need. Advances in neural information processing systems  \textbf{30} (2017)

\bibitem{Wang2017}
Wang, F., Jiang, M., Qian, C., Yang, S., Li, C., Zhang, H., Wang, X., Tang, X.: Residual attention network for image classification. In: IEEE Conference on Computer Vision and Pattern Recognition (CVPR). pp. 6450--6458 (2017)

\bibitem{Wei2022}
Wei, X., Dai, L.: Channel estimation for extremely large-scale massive mimo: Far-field, near-field, or hybrid-field? IEEE Communications Letters  \textbf{26}(1),  177--181 (2022)

\bibitem{Zhang2023}
Zhang, X., Wang, Z., Zhang, H., Yang, L.: Near-field channel estimation for extremely large-scale array communications: A model-based deep learning approach. IEEE Communications Letters  \textbf{27}(4),  1155--1159 (2023)

\end{thebibliography}

\end{document}